\documentclass[12pt]{article}
\usepackage[margin=1.9cm]{geometry}
\usepackage{graphicx}
\usepackage{cite}
\usepackage{xcolor}
\usepackage{feynmf}

\newcommand{\mysection}{\setcounter{equation}{0}\section}

\def\beq{\begin{equation}}
\def\eeq{\end{equation}}
\def\beqa{\begin{eqnarray}}
\def\eeqa{\end{eqnarray}}
 
\begin{document}

\begin{center}
{\Large \bf N$^3$LO soft-gluon corrections in single-particle-inclusive kinematics and $H^+ H^-$ production}
\end{center}

\vspace{2mm}

\begin{center}
{\large Nikolaos Kidonakis and Alberto Tonero}\\

\vspace{2mm}

{\it Department of Physics, Kennesaw State University, \\
Kennesaw, GA 30144, USA}

\end{center}

\begin{abstract}
We calculate the complete soft-gluon corrections for the production of colorless final states through N$^3$LO in single-particle-inclusive kinematics. We present explicit analytical results and use them to study higher-order QCD corrections for the production of a heavy charged Higgs pair ($H^+ H^-$) via quark-antiquark annihilation in the Two-Higgs-Doublet Model at LHC energies. We calculate the NNLO soft-gluon and virtual QCD corrections as well as the  N$^3$LO soft-gluon corrections to the total cross section and the charged-Higgs rapidity distribution. This is the first calculation of complete N$^3$LO soft-gluon corrections for a process in single-particle-inclusive kinematics, and the results can be applied to other processes with colorless final states.
\end{abstract}

\mysection{Introduction}

Precision in calculations of cross sections for hard-scattering processes requires the inclusion of higher-order QCD corrections. A particular class of these corrections arises from the emission of soft gluons. These soft radiative corrections can be formally resummed \cite{GS87,CT89,NKGS1,NKGS2,LOS} to all orders in the strong coupling after taking Mellin moments or Laplace transforms of the cross section and using its factorization properties, and the resummed cross section can be used as a generator of fixed-order corrections. Soft-gluon resummation has been successfully developed and applied to processes with massive final states, such as top-quark pair production \cite{NKGS1,NKGS2,LOS,NK2loop,NKnnll,NKan3lo}, because the cross section receives large contributions from soft-gluon emission near partonic threshold due to the large mass. In all these processes, the soft-gluon corrections are numerically dominant and account for the overwhelming majority of the complete corrections at higher orders. 

For processes with complex color flows, such as top production, one does not yet have all the ingredients necessary to calculate the complete set of soft-gluon corrections at next-to-next-to-next-to-leading order (N$^3$LO). For processes with trivial color flow, and in particular with colorless (e.g. electroweak) final states, the N$^3$LO soft-gluon corrections are known in fully inclusive kinematics \cite{MA05,VR06,NKhiggs}. However, complete N$^3$LO soft-gluon contributions have not yet been calculated for a partonic process in single-particle-inclusive (1PI) kinematics. In this paper we provide explicit analytical results for the complete soft-gluon corrections through N$^3$LO for 1PI partonic processes with no color in the final state. As a concrete application, we employ these results to compute higher-order QCD corrections for charged-Higgs pair production in the Two-Higgs-Doublet Model (2HDM) using 1PI kinematics. 

In this work we present the most precise predictions for quark-induced production of $H^+H^-$ at the LHC, that occurs through diagrams of Drell-Yan type, in the 2HDM by computing the total cross section and the charged-Higgs rapidity distribution for this process at approximate next-to-next-to-leading order (aNNLO) and approximate N$^3$LO (aN$^3$LO) in QCD. Results at aNNLO  are derived by adding the second-order soft-plus-virtual corrections to the full next-to-leading order (NLO) QCD calculation, while aN$^3$LO results are obtained by further adding the third-order soft-gluon corrections.

The 2HDM represents one of the simplest extensions of the Standard Model Higgs sector, introducing an additional $SU(2)_L$ Higgs doublet.  This model gives rise to a very rich phenomenology due to its extended Higgs sector (for a review see~\cite{2HDMreview} and references therein). The model predicts the existence of three neutral scalar bosons ($h$, $H$, and $A$) and a pair of charged Higgs bosons ($H^+$ and $H^-$).

The dominant production channel for the charged Higgs at the Large Hadron Collider (LHC) is in association with a top quark $p\,p\to t\,H^-$, for which soft-gluon corrections were calculated in Refs. \cite{NKtWH,NKcH,NKcHnnlo}, followed by the associated production with a $W$ boson $p\,p\to H^+W^-$ (see \cite{NKWH} for soft-gluon corrections), and pair production $p\,p\to H^+H^-$. In this work we focus on the last process, which can occur at leading order (LO) through quark-antiquark annihilation ($q\bar q \to H^+H^-$) \cite{Djouadi:1999ht,Hou:2005alu,Alves:2005kr} and gluon fusion ($gg \to H^+H^-$) \cite{Jiang:1997cg,Krause:1997rc,Jiang:1997cr,BarrientosBendezu:1999gp,Brein:1999sy}. In the case of light quarks in the initial state, the pair production of charged Higgs bosons is of Drell-Yan type, with the exchange of a photon and a $Z$ boson in the $s$-channel. NLO QCD corrections for this channel have been computed in the 2HDM and the Minimal Supersymmetric Standard Model (MSSM) in~\cite{Djouadi:1999ht,Alves:2005kr}. In the case of bottom quarks in the initial states, there are additional diagrams that account for the exchange of a neutral Higgs boson in the $s$-channel and a top quark in the $t$-channel. The NLO QCD corrections to $b\bar b \to H^+H^-$ have been computed in~\cite{Alves:2005kr,Hou:2005alu}. The charged-Higgs pair production through gluon fusion is a one-loop process already at LO and it has been computed in the 2HDM and the MSSM in~\cite{Jiang:1997cg,Krause:1997rc,Jiang:1997cr,BarrientosBendezu:1999gp,Brein:1999sy}. 

In models like the type II 2HDM, the quark-antiquark annihilation process $q\bar q\to H^+H^-$ is the dominant channel for low values of $\tan\beta$, which is the ratio of the vacuum expectation values for the two doublets. On the other hand, for large values of $\tan\beta$, the $gg$ and $b\bar b$ channels have cross sections which are enhanced by a $\tan^4\beta$ factor, and this can compensate for the loop suppression or the small bottom-quark parton densities and can make these channels dominant, as shown in~\cite{Alves:2005kr}. However, recent studies showed that, for type II 2DHM with softly broken $Z_2$ symmetry, the size of $\cos(\beta-\alpha)$ has to be lower than ${\cal O}(10^{-2})$ and that low values of $\tan\beta$ are favored~\cite{Chowdhury:2017aav,Atkinson:2021eox,Wang:2022yhm,Arco:2022jrt}. Therefore, the calculations presented in this work are particularly relevant for those scenarios.

The paper is organized as follows. In Section 2, we describe the soft-gluon resummation formalism and provide explicit analytical results for the soft-gluon corrections through N$^3$LO in 1PI kinematics for processes with colorless final states. In Section 3 we introduce the Two-Higgs-Doublet model and apply the resummation formalism of Section 2 to charged-Higgs pair production. In Section 4, we present numerical results for the total cross section through aN$^3$LO at LHC energies. In Section 5, we present results for the charged-Higgs rapidity distribution for two representative masses of the charged Higgs boson. We conclude in Section 6.

\mysection{Soft-gluon corrections at N$^3$LO in 1PI kinematics}

In this section, we discuss the resummation formalism that we use for the calculation of soft-gluon corrections in 1PI kinematics for partonic processes with colorless final states, e.g. with electroweak final-state particles, such as $H^+ H^-$ production. The origin of the soft-gluon corrections is from the emission of soft (i.e. low-energy) gluons from the initial state partons, which result in partial cancellations of infrared divergences between real-emission and virtual diagrams close to partonic threshold. We derive the resummed cross section via Laplace transforms, factorization, and renormalization-group evolution. We then provide explicit analytical results for the soft-gluon corrections through N$^3$LO.

We consider first leading-order partonic processes of the form 
\beq
q (p_a)\, + {\bar q} (p_b) \to A (p_1) \, + \, B (p_2)  \, ,
\eeq
where $q $ and ${\bar q}$ are quarks and antiquarks in the protons, and $A$ and $B$ are a pair of colorless particles, each of mass $m$ 
(e.g. $H^+H^-$), with particle $A$ the observed particle in 1PI kinematics. We will also discuss later the changes needed in the analytical expressions if we instead have gluons in the initial state, i.e. processes $gg \to AB$, or if the two final-state particles have different masses, $m_A$ and $m_B$. We define $s=(p_a+p_b)^2$, $t_1=(p_a-p_1)^2-m^2$, $u_1=(p_b-p_1)^2-m^2$, as well as a partonic threshold variable $s_4=s+t_1+u_1=(p_2+p_g)^2-m^2$, with $p_g$ the momentum of an additional gluon in the final state. Near partonic threshold $p_g \to 0$ and, thus, $s_4 \to 0$. 
The soft-gluon corrections appear in the perturbative series as logarithms of $s_4$, i.e. $[(\ln^k(s_4/m^2))/s_4]_+$, with $0 \le k \le 2n-1$ at $n$th order in the strong coupling, $\alpha_s$.

The resummation of soft-gluon corrections is a consequence of the factorization properties of the (in general, differential) cross section under Laplace transforms in 1PI kinematics. We first write the differential hadronic cross section, $d\sigma_{pp \to AB}$, as a convolution of the differential partonic cross section, $d{\hat \sigma}_{q{\bar q} \to AB}$, with the parton distribution functions (pdf), $\phi_{q/p}$ and $\phi_{{\bar q}/p}$, as 
\beq
d\sigma_{pp \to AB}=\sum_{q,{\bar q}} \; 
\int dx_a \, dx_b \,  \phi_{q/p}(x_a, \mu_F) \, \phi_{{\bar q}/p}(x_b, \mu_F)  \, 
d{\hat \sigma}_{q{\bar q} \to AB}(s_4, \mu_F)  \, ,
\label{factorized}
\eeq
where $\mu_F$ is the factorization scale, and $x_a$, $x_b$ are momentum fractions of partons $q$, ${\bar q}$, respectively, in the colliding protons. 
We also define the hadron-level variables $S=(P_a+P_b)^2$, $T_1=(P_a-p_1)^2-m^2$, $U_1=(P_b-p_1)^2-m^2$, where $P_a$ and $P_b$ denote the momenta of the colliding protons, and $S_4=S+T_1+U_1$. Since $p_a=x_a P_a$ and $p_b=x_b P_b$, we have the relations $s=x_a x_b S$, $t_1=x_a T_1$, $u_1=x_b U_1$, and
\beq
\frac{S_4}{S}=\frac{s_4}{s}-(1-x_a) \frac{u_1}{s}-(1-x_b) \frac{t_1}{s} \, .
\label{S4}
\eeq

We then consider the parton-parton cross section $d\sigma_{q{\bar q} \to AB}$, which is of the same form as Eq.~(\ref{factorized}) but with incoming partons instead of hadrons \cite{NKGS1,NKGS2,LOS},  
\beq
d\sigma_{q{\bar q} \to AB}(S_4)=
\int dx_a \, dx_b \,  \phi_{q/q}(x_a) \, \phi_{{\bar q}/{\bar q}}(x_b) \, 
d{\hat \sigma}_{q{\bar q} \to AB}(s_4) \, ,
\label{factphi}
\eeq
and we define its Laplace transform as 
\beq
d{\tilde \sigma}_{q{\bar q} \to AB}(N)=\int_0^S 
\frac{dS_4}{S} \,  e^{-N S_4/S} \, d\sigma_{q{\bar q} \to AB}(S_4) \, . 
\label{cstr}
\eeq 

Using the expression for $S_4/S$ in Eq. (\ref{S4}), we can rewrite the Laplace transform, Eq. (\ref{cstr}), of the parton-parton cross section in Eq. (\ref{factphi}) as  
\beqa
d{\tilde \sigma}_{q{\bar q} \to AB}(N) &=& \int_0^1 dx_a e^{-N_a (1-x_a)} 
\phi_{q/q}(x_a) \int_0^1 dx_b e^{-N_b (1-x_b)} \phi_{{\bar q}/{\bar q}}(x_b)
\int_0^s \frac{ds_4}{s} e^{-N s_4/s} d{\hat \sigma}_{q{\bar q} \to AB}(s_4)
\nonumber \\ 
&=& {\tilde \phi}_{q/q}(N_a) \, {\tilde \phi}_{{\bar q}/{\bar q}}(N_b) \, 
d{\tilde{\hat \sigma}}_{q{\bar q} \to AB}(N) \, ,
\label{fac}
\eeqa
where $N_a=N(-u_1/s)$ and $N_b=N(-t_1/s)$.

Next, we introduce a refactorization of the cross section via new functions $H_{q{\bar q} \to AB}$, $S_{q{\bar q} \to AB}$, $\psi_{q/q}$, $\psi_{{\bar q}/{\bar q}}$ \cite{NKGS1,NKGS2,LOS}. The hard function $H_{q{\bar q} \to AB}$ is purely short-distance and infrared safe while the soft function $S_{q{\bar q} \to AB}$ describes the emission of noncollinear soft gluons. The coupling of the soft gluons to the partons in the hard-scattering process is described by eikonal (Wilson) lines, i.e. ordered exponentials of the gauge field. The functions $\psi_{q/q}$ and $\psi_{{\bar q}/{\bar q}}$ differ from the pdf $\phi_{q/q}$ and $\phi_{{\bar q}/{\bar q}}$, and they describe collinear emission from the incoming partons \cite{GS87,NKGS1,NKGS2,LOS}. The refactorized form of the cross section \cite{LOS,NKVD} is then 
\beqa
d{\sigma}_{q{\bar q} \to AB}&=&\int dw_a \, dw_b \, dw_S \, \psi_{q/q}(w_a) \, \psi_{{\bar q}/{\bar q}}(w_b) 
\nonumber \\ && \times
H_{q{\bar q} \to AB} \, \, 
S_{q{\bar q} \to AB}\left(\frac{w_S \sqrt{s}}{\mu_F} \right) \; 
\delta\left(\frac{S_4}{S}-w_S+w_a\frac{u_1}{s}
+w_b \frac{t_1}{s}\right)
\label{refact}
\eeqa
where the $w$'s are dimensionless weights, with $w_a$ and $w_b$ for $\psi_{q/q}$ and $\psi_{{\bar q}/{\bar q}}$, respectively, and $w_S$ for $S_{q{\bar q} \to AB}$. The argument in the delta function of Eq. (\ref{refact}) arises from rewriting Eq. (\ref{S4}) in terms of the new weights \cite{LOS}, as
\beq
\frac{S_4}{S}=w_S-w_a \frac{u_1}{s}- w_b \frac{t_1}{s}\, .
\label{ws}
\eeq

After taking a Laplace transform of Eq. (\ref{refact}), we have
\beqa
d{\tilde \sigma}_{q{\bar q} \to AB}(N)&=& 
\int_0^1 dw_a \, e^{-N_a w_a} \, \psi_{q/q}(w_a) \int_0^1 dw_b \, e^{-N_b w_b} \, \psi_{{\bar q}/{\bar q}}(w_b)
\nonumber \\ && \times \, \,
H_{q{\bar q} \to AB}  \int_0^1 dw_s \, e^{-N w_s} \, S_{q{\bar q} \to AB}\left(\frac{w_s\sqrt{s}}{\mu_F} \right)
\nonumber \\ &=& 
{\tilde \psi}_{q/q}(N_a) \, \, {\tilde \psi}_{{\bar q}/{\bar q}}(N_b) \, \, 
H_{q{\bar q} \to AB} \, \, 
{\tilde S}_{q{\bar q} \to AB}\left(\frac{\sqrt{s}}{N \mu_F} \right) \, .
\label{refac}
\eeqa
All $N$-dependence has now been absorbed into the functions ${\tilde \psi}_{q/q}$, ${\tilde \psi}_{{\bar q}/{\bar q}}$, and ${\tilde S}$.

Comparing Eqs. (\ref{fac}) and (\ref{refac}), we get an expression for the hard-scattering partonic cross section in Laplace transform space,
\beq
d{\tilde{\hat \sigma}}_{q{\bar q} \to AB}(N)=
\frac{{\tilde \psi}_{q/q}(N_a) \, {\tilde \psi}_{{\bar q}/{\bar q}}(N_b)}
{{\tilde \phi}_{q/q}(N_a) \, {\tilde \phi}_{{\bar q}/{\bar q}}(N_b)} \, \,  
H_{q{\bar q} \to AB} \, \, 
{\tilde S}_{q{\bar q} \to AB}\left(\frac{\sqrt{s}}{N \mu_F} \right) \, .
\label{sigN}
\eeq

The resummed differential cross section in Laplace-transform space is derived from the renormalization-group evolution of ${\tilde \psi}_{q/q}/{\tilde \phi}_{q/q}$, ${\tilde \psi}_{{\bar q}/{\bar q}}/{\tilde \phi}_{{\bar q}/{\bar q}}$, and ${\tilde S}_{q{\bar q} \to AB}$ in Eq. (\ref{sigN}), and it is given by the expression \cite{GS87,CT89,NKGS1,NKGS2,LOS}
\beqa
d{\tilde{\hat \sigma}}^{res}_{q{\bar q} \to AB}(N)&=&   
\exp\left[E_q (N_a)+ E_{\bar q} (N_b)\right] \; \,
\exp \left[2\int_{\mu_F}^{\sqrt{s}} \frac{d\mu}{\mu}\; 
\left(\gamma_{q/q}\left(N_a,\alpha_s(\mu)\right)
+\gamma_{{\bar q}/{\bar q}}\left(N_b,\alpha_s(\mu)\right)\right)\right] \nonumber \\ && \times \, \, 
H_{q{\bar q} \to AB}\left(\alpha_s({\sqrt s})\right) \, \,
{\tilde S}_{q{\bar q} \to AB}\left(\alpha_s({\sqrt s/N})\right) \, .
\label{resum}
\eeqa
In the first exponential of Eq. (\ref{resum}), we have
\beq
E_q(N_a)=
\int^1_0 dz \frac{z^{N_a-1}-1}{1-z}\;
\left \{\int_1^{(1-z)^2} \frac{d\lambda}{\lambda}
A_q \left(\alpha_s(\lambda s)\right)
+D_q \left[\alpha_s((1-z)^2 s)\right]\right\} \, ,
\label{Eexp}
\eeq
with an analogous expression for $E_{\bar q}(N_b)$ \cite{GS87}. 
The first integrand in Eq.~(\ref{Eexp}) has the perturbative expansion 
$A_q = (\alpha_s/\pi) A_q^{(1)}+(\alpha_s/\pi)^2 A_q^{(2)}+(\alpha_s/\pi)^3 A_q^{(3)}+\cdots$.
The first term in this expansion is $A_q^{(1)}=C_F$ \cite{GS87}, where $C_F=(N_c^2-1)/(2N_c)$ and $N_c=3$ is the number of colors. The second term, $A_q^{(2)}$, is given by \cite{CT89,KT82}
\beq
A_q^{(2)} =C_F C_A \left(\frac{67}{36}-\frac{\zeta_2}{2}\right)-\frac{5}{18}C_F n_f\,,
\eeq         
where $C_A=N_c$ and $n_f$ is the number of light-quark flavors (in the next section we will set $n_f=5$ for $H^+H^-$ production). Finally, the third term, $A_q^{(3)}$, is given by \cite{MVV04}
\beqa
A_q^{(3)}&=&C_F C_A^2\left(\frac{245}{96}-\frac{67}{36}\zeta_2
+\frac{11}{24}\zeta_3+\frac{11}{8}\zeta_4\right)
+C_F^2 n_f\left(-\frac{55}{96}+\frac{\zeta_3}{2}\right)
\nonumber \\ &&
{}+C_F C_A n_f \left(-\frac{209}{432}+\frac{5}{18}\zeta_2
-\frac{7}{12} \zeta_3\right)-C_F\frac{n_f^2}{108} \, .
\eeqa
Here and below $\zeta_2=\pi^2/6$, $\zeta_3=1.202056903\cdots$,  
$\zeta_4=\pi^4/90$, and $\zeta_5=1.036927755\cdots$.  
The corresponding expressions for processes with gluons in the initial state (i.e. $gg \to AB$) are given at these perturbative orders by
$A_g^{(n)}=(C_A/C_F) A_q^{(n)}$, where $n=1,2,3$.

The second integrand in Eq.~(\ref{Eexp}) can be expanded as $D_q=(\alpha_s/\pi)D_q^{(1)}+(\alpha_s/\pi)^2 D_q^{(2)}
+(\alpha_s/\pi)^3 D_q^{(3)}+\cdots$. In Feynman gauge, the one-loop term $D_q^{(1)}=0$, while
at two loops \cite{CLS97}
\beq
D_q^{(2)}=C_F C_A \left(-\frac{101}{54}+\frac{11}{6} \zeta_2
+\frac{7}{4}\zeta_3\right)
+C_F n_f \left(\frac{7}{27}-\frac{\zeta_2}{3}\right) \, ,
\eeq
and at three loops \cite{MA05} 
\beqa
D_q^{(3)}&=&C_F C_A^2 \left(-\frac{297029}{46656}
+\frac{6139}{648} \zeta_2+\frac{2509}{216} \zeta_3
-\frac{187}{48} \zeta_4 -\frac{11}{12} \zeta_2 \zeta_3-3 \zeta_5\right)
\nonumber \\ && 
{}+C_F C_A n_f \left(\frac{31313}{23328}-\frac{1837}{648}\zeta_2
-\frac{155}{72}\zeta_3+\frac{23}{24}\zeta_4\right)
\nonumber \\ && 
{}+C_F^2 n_f \left(\frac{1711}{1728}-\frac{\zeta_2}{4}-\frac{19}{36}\zeta_3
-\frac{\zeta_4}{4}\right)
+C_F n_f^2 \left(-\frac{29}{729}+\frac{5}{27}\zeta_2
+\frac{5}{54}\zeta_3 \right) \, .
\eeqa
The corresponding expressions for processes with gluons in the initial state are given at these perturbative orders by
$D_g^{(n)}=(C_A/C_F) D_q^{(n)}$, where $n=1,2,3$.

In the second exponent of Eq.~(\ref{resum}), the quantity $\gamma_{q/q}$ is the moment-space 
anomalous dimension of the ${\overline {\rm MS}}$ density $\phi_{q/q}$
\cite{FRS1,FRS2,GALY79,GFP80,FP80} which can be expressed as $\gamma_{q/q}(N)=-A_q \ln {\tilde N} +\gamma_q$, and similarly for $\gamma_{{\bar q}/{\bar q}}$. The parton anomalous dimension can be expanded as
$\gamma_q=(\alpha_s/\pi) \gamma_q^{(1)}
+(\alpha_s/\pi)^2 \gamma_q^{(2)} + \cdots$
with $\gamma_q^{(1)}=3C_F/4$ and
\beq
\gamma_q^{(2)}=C_F^2\left(\frac{3}{32}-\frac{3}{4}\zeta_2
+\frac{3}{2}\zeta_3\right)
+C_F C_A\left(\frac{17}{96}+\frac{11}{12}\zeta_2-\frac{3}{4}\zeta_3\right)
-C_F n_f \left(\frac{1}{48}+\frac{\zeta_2}{6}\right)\, .
\eeq
For gluons, we have $\gamma_g^{(1)}=\beta_0/4$, where $\beta_0=11 C_A/3-2n_f/3$ is the coefficient of the first term in the perturbative expansion of the $\beta$ function \cite{GW,HDP}, and
\beq
\gamma_g^{(2)}=C_A^2\left(\frac{2}{3}+\frac{3}{4}\zeta_3\right)
-n_f \left(\frac{C_F}{8}+\frac{C_A}{6}\right)\, .
\eeq

By expanding Eq. (\ref{resum}) to fixed order and inverting back to momentum space, we can calculate the soft-gluon corrections completely through N$^3$LO.
The Born differential cross section at partonic level can be written as
\beq
\frac{d^2{\hat{\sigma}}^{(0)}_{q{\bar q} \to AB}}{dt_1 \, du_1} = F^B_{q{\bar q} \to AB} \, \delta(s_4) \, ,
\label{LO}
\eeq
where the specific form of $F^B_{q{\bar q} \to AB}$ depends on the process (see the next section for $H^+H^-$ production). If we perturbatively expand $H_{q{\bar q} \to AB}=H_{q{\bar q} \to AB}^{(0)}+(\alpha_s/\pi) H_{q{\bar q} \to AB}^{(1)}+(\alpha_s/\pi)^2 H_{q{\bar q} \to AB}^{(2)}+\cdots$, and $S_{q{\bar q} \to AB}=S_{q{\bar q} \to AB}^{(0)}+(\alpha_s/\pi) S_{q{\bar q} \to AB}^{(1)}+(\alpha_s/\pi)^2 S_{q{\bar q} \to AB}^{(2)}+\cdots$, then we have $F^B_{q{\bar q} \to AB}=H_{q{\bar q} \to AB}^{(0)} \, S_{q{\bar q} \to AB}^{(0)}$.

The NLO soft-plus-virtual corrections are given by
\beq
\frac{d^2{\hat{\sigma}}^{(1)}_{q{\bar q} \to AB}}{dt_1 \, du_1} = F^B_{q{\bar q} \to AB}
\frac{\alpha_s(\mu_R)}{\pi}
\left\{c_3\, {\cal D}_1(s_4) + c_2\,  {\cal D}_0(s_4) 
+c_1\,  \delta(s_4)\right\}
\label{NLO}
\eeq
where $\mu_R$ is the renormalization scale and
\beq
{\cal D}_k(s_4)=\left[\frac{\ln^k(s_4/m^2)}{s_4}\right]_+\,.
\label{Dplus}
\eeq
The coefficients of the ${\cal D}_1(s_4)$ and ${\cal D}_0(s_4)$ terms in Eq. (\ref{NLO}) are, respectively, 
\beq
c_3=4 C_F\qquad{\rm and}\qquad
c_2= -2 \, C_F \, \ln\left(\frac{t_1 u_1}{m^4}\right)
-2 \, C_F \, \ln\left(\frac{\mu_F^2}{s}\right) \, .
\label{c3c2}
\eeq
The coefficient of the $\delta(s_4)$ term is
\beq
c_1= C_F\, \ln^2\left(\frac{-t_1}{m^2}\right)
+ C_F\, \ln^2\left(\frac{-u_1}{m^2}\right)
+\left[C_F \, \ln\left(\frac{t_1 u_1}{m^4}\right)-2\gamma_q^{(1)}\right]
\ln\left(\frac{\mu_F^2}{s}\right)+V_1 \, ,
\label{c1}
\eeq
where $V_1=\left(H_{q{\bar q} \to AB}^{(1)} \, S_{q{\bar q} \to AB}^{(0)}+H_{q{\bar q} \to AB}^{(0)} \, S_{q{\bar q} \to AB}^{(1)}\right)/ F^B_{q{\bar q} \to AB}$ is the contribution from the one-loop virtual corrections. For Drell-Yan type processes with quark-antiquark annihilation, such as $H^+H^-$ production, $V_1$ is given by
the expression $V_1=2C_F(-2+\zeta_2)$ \cite{KAP,AEM,HvN}. For gluon-initiated processes, we replace $C_F$ by $C_A$ in Eqs. (\ref{c3c2}) and (\ref{c1}), and we use the corresponding $V_1$ (for example, see \cite{DSZ,SD} for the NLO virtual corrections for $gg\to H$).

The NNLO soft-plus-virtual corrections are given by
\beqa
\frac{d^2{\hat{\sigma}}^{(2)}_{q{\bar q} \to AB}}{dt_1 \ du_1}&=&F^B_{q{\bar q} \to AB} \frac{\alpha_s^2(\mu_R)}{\pi^2}
\left\{\frac{1}{2}c_3^2\, {\cal D}_3(s_4) + 
\left[\frac{3}{2}c_3 c_2-\frac{\beta_0}{4} c_3 \right]  {\cal D}_2(s_4) \right.
\nonumber \\ &&  \hspace{-5mm}
{}+\left[c_3 c_1+c_2^2-\zeta_2 c_3^2
-\frac{\beta_0}{2} c_2-\beta_0 C_F
\ln\left(\frac{\mu_F^2}{\mu_R^2}\right)+4 A_q^{(2)}\right] {\cal D}_1(s_4)
\nonumber \\ &&    \hspace{-5mm}
{}+\left[c_2 c_1-\zeta_2 c_3 c_2+\zeta_3 c_3^2
+\frac{\beta_0}{4} c_2 \ln\left(\frac{\mu_R^2}{s}\right) \right. 
\nonumber \\ && \;
{}-\frac{\beta_0}{2} C_F \ln^2\left(\frac{-t_1}{m^2}\right)
-\frac{\beta_0}{2} C_F \ln^2\left(\frac{-u_1}{m^2}\right) 
-2 A_q^{(2)} \ln\left(\frac{t_1 u_1}{m^4}\right)+2 D_q^{(2)}
\nonumber \\ &&  \; \left. 
{}+\frac{\beta_0}{4} C_F \ln^2\left(\frac{\mu_F^2}{s}\right) 
-2 A_q^{(2)} \ln\left(\frac{\mu_F^2}{s}\right)\right]
{\cal D}_0(s_4)
\nonumber \\ &&   \hspace{-5mm}
{}+\left[V_2+\frac{1}{2} \left(c_1^2-V_1^2\right)-\frac{\zeta_2}{2}c_2^2
+\zeta_3 c_3 c_2 +\frac{\beta_0}{6} C_F \left(\ln^3\left(\frac{-t_1}{m^2}\right)+\ln^3\left(\frac{-u_1}{m^2}\right)\right) \right.
\nonumber \\ && \;
{}+\left(\frac{\beta_0}{4} C_F+A_q^{(2)}\right) \left(\ln^2\left(\frac{-t_1}{m^2}\right)+\ln^2\left(\frac{-u_1}{m^2}\right)\right)
+\frac{\beta_0}{4} c_1 \ln\left(\frac{\mu_R^2}{s}\right)
-2 \gamma_q^{(2)} \ln\left(\frac{\mu_F^2}{s}\right)
\nonumber \\ && \; \left. \left. 
{}+A_q^{(2)} \ln\left(\frac{t_1 u_1}{m^4}\right) \ln\left(\frac{\mu_F^2}{s}\right)
+\frac{\beta_0}{8} \left(2 \gamma_q^{(1)}-C_F \ln\left(\frac{t_1 u_1}{m^4}\right)\right)
\ln^2\left(\frac{\mu_F^2}{s}\right) \right] \delta(s_4) \right\} 
\nonumber \\
\label{NNLO}
\eeqa
where $V_2=\left(H_{q{\bar q} \to AB}^{(2)} \, S_{q{\bar q} \to AB}^{(0)}+H_{q{\bar q} \to AB}^{(0)} \, S_{q{\bar q} \to AB}^{(2)}+H_{q{\bar q} \to AB}^{(1)} \, S_{q{\bar q} \to AB}^{(1)}\right)/F^B_{q{\bar q} \to AB}$ is the contribution from the virtual two-loop corrections.
For Drell-Yan type processes with quark-antiquark annihilation, such as $H^+H^-$ production, $V_2$ is given by the expression \cite{V2qq}
\beqa
V_2&=&C_F^2\left(\frac{511}{64}-\frac{35}{8}\zeta_2-\frac{15}{4}\zeta_3
+\frac{\zeta_2^2}{10}\right)+C_F C_A\left(-\frac{1535}{192}
+\frac{37}{9}\zeta_2+\frac{7}{4}\zeta_3-\frac{3}{20}\zeta_2^2\right)
\nonumber \\ &&
{}+C_F n_f \left(\frac{127}{96}-\frac{7}{9}\zeta_2+\frac{\zeta_3}{2}\right) \, .
\eeqa
For processes $gg \to AB$, we replace in Eq. (\ref{NNLO}) all explicit appearances of $C_F$ by $C_A$, and we also use the corresponding expressions for $A_g$, $D_g$, $\gamma_g$, $c_3$, $c_2$, and $c_1$ as detailed above, as well as the appropriate expressions for $V_1$ and $V_2$ (for example, see \cite{RVH} for the NNLO virtual corrections for $gg\to H$).

The N$^3$LO soft-gluon corrections are given by
\beqa
\frac{d^2{\hat{\sigma}}^{(3)}_{q{\bar q} \to AB}}{dt_1 \, du_1}&=& 
F^B_{q{\bar q} \to AB} \frac{\alpha_s^3(\mu_R)}{\pi^3} \left\{  
\frac{1}{8} c_3^3 \; {\cal D}_5(s_4)
+\left[\frac{5}{8} c_3^2 c_2 -\frac{5}{24} c_3^2 \beta_0 \right] \;  {\cal D}_4(s_4) \right.
\nonumber \\ && \hspace{-5mm}
{}+\left[c_3 c_2^2 +\frac{1}{2}c_3^2 c_1-\zeta_2 c_3^3 
+\frac{\beta_0^2}{12}c_3-\frac{5}{6}\beta_0 c_3 c_2
-\beta_0 C_F c_3 \ln\left(\frac{\mu_F^2}{\mu_R^2}\right) 
+4 c_3 A_q^{(2)} \right] \; {\cal D}_3(s_4)
\nonumber \\ &&  \hspace{-5mm}
{}+\left[\frac{3}{2} c_3 c_2 c_1 +\frac{1}{2} c_2^3
-3 \zeta_2 c_3^2 c_2 +\frac{5}{2} \zeta_3 c_3^3
-\frac{\beta_0}{4} c_3 c_1 +\frac{9}{8} \beta_0 \zeta_2 c_3^2 \right.
\nonumber \\ && \; \left.
{}+\left(3 c_2-\beta_0 \right) \left(-\frac{\beta_0}{4} c_2
-\frac{\beta_0}{2} C_F \ln\left(\frac{\mu_F^2}{\mu_R^2}\right) +2 A_q^{(2)} \right)
-C_F \frac{\beta_1}{4}-\frac{3}{2} c_3 X_1 \right] {\cal D}_2(s_4)
\nonumber \\ &&  \hspace{-5mm}
{}+\left[\frac{1}{2} c_3 c_1^2+c_2^2 c_1 -\zeta_2 c_3^2 c_1
-\frac{5}{2} \zeta_2 c_3 c_2^2+5 \zeta_3 c_3^2 c_2+\frac{5}{4} \zeta_2^2 c_3^3
-\frac{15}{4} \zeta_4 c_3^3 \right.
\nonumber \\ && \;        
{}-\frac{\beta_0^2}{4} \zeta_2 c_3
-\frac{5}{3} \beta_0 \zeta_3 c_3^2+\beta_0 \zeta_2 c_3 c_2
+\left(2 c_1-5 \zeta_2 c_3\right) \left(-\frac{\beta_0}{4}c_2
-\frac{\beta_0}{2} C_F \ln\left(\frac{\mu_F^2}{\mu_R^2}\right)+2 A_q^{(2)}\right)
\nonumber \\ && \;
{}+(\beta_0 -2 c_2) X_1+c_3 X_0 +4 A_q^{(3)}
+C_F \frac{\beta_0^2}{4} \ln^2\left(\frac{\mu_F^2}{\mu_R^2}\right)
-2\beta_0 A_q^{(2)} \ln\left(\frac{\mu_F^2}{\mu_R^2}\right)
\nonumber \\ && \; \left.
{}+C_F \frac{\beta_1}{4} \ln\left(\frac{\mu_R^2}{s}\right)
+C_F \frac{\beta_1}{4}\ln\left(\frac{t_1 u_1}{m^4}\right) \right] \;
{\cal D}_1(s_4)
\nonumber \\ &&  \hspace{-5mm}
{}+\left[\frac{1}{2} c_2 c_1^2+3 \zeta_5 c_3^3-\frac{15}{4} \zeta_4 c_3^2 c_2
-2 \zeta_2 \zeta_3 c_3^3+ \zeta_3 c_3^2 c_1+2 \zeta_3 c_3 c_2^2
+\frac{5}{4} \zeta_2^2 c_3^2 c_2 -\zeta_2 c_3 c_2 c_1-\frac{\zeta_2}{2} c_2^3
\right.
\nonumber \\ && \;
{}+\frac{\beta_0}{12} c_3 \left(15 \zeta_4 c_3-8 \zeta_3 c_2-6 \zeta_2^2 c_3+3 \zeta_2 c_1\right)
\nonumber \\ && \; 
{}+\left(4 \zeta_3 c_3-3 \zeta_2 c_2\right) \left(-\frac{\beta_0}{4} c_2
-\frac{\beta_0}{2} C_F \ln\left(\frac{\mu_F^2}{\mu_R^2}\right)+2 A_q^{(2)}\right)
\nonumber \\ && \; 
{}+\left(\zeta_2 c_3 -c_1 \right) X_1 + c_2 X_0
-\frac{\beta_0^2}{4} C_F \left(\ln^2\left(\frac{-t_1}{m^2}\right)
+\ln^2\left(\frac{-u_1}{m^2}\right)\right) \ln\left(\frac{\mu_R^2}{s}\right)
\nonumber \\ && \; 
{}+\frac{\beta_0^2}{16} c_2 \ln^2\left(\frac{\mu_R^2}{s}\right)
+\frac{\beta_0^2}{8} C_F  \ln\left(\frac{\mu_F^2}{s}\right) \ln^2\left(\frac{\mu_R^2}{s}\right)
\nonumber \\ && \; \left. \left.
{}-\frac{\beta_1}{8} C_F \left(\ln^2\left(\frac{-t_1}{m^2}\right)
+\ln^2\left(\frac{-u_1}{m^2}\right)\right)+Y_0 \right] \;
{\cal D}_0(s_4) \right\}
\label{N3LO}
\eeqa
where
\beqa
X_1&=&\frac{\beta_0}{4} \zeta_2 c_3 
-\frac{\beta_0}{4} c_2 \ln\left(\frac{\mu_R^2}{s}\right)
+\frac{\beta_0}{2} C_F \ln^2\left(\frac{-t_1}{m^2}\right)
+\frac{\beta_0}{2} C_F \ln^2\left(\frac{-u_1}{m^2}\right)
\nonumber \\ &&
{}+2 A_q^{(2)} \ln\left(\frac{t_1 u_1}{m^4}\right)-2D_q^{(2)}  
-\frac{\beta_0}{4} C_F \ln^2\left(\frac{\mu_F^2}{s}\right)
+2 A_q^{(2)} \ln\left(\frac{\mu_F^2}{s}\right) \, , 
\label{X1}
\eeqa
\beqa
X_0&=& V_2-\frac{1}{2}V_1^2 -\frac{1}{4} \zeta_2^2 c_3^2 +\frac{3}{4} \zeta_4 c_3^2
+\frac{\beta_0}{4} c_1 \ln\left(\frac{\mu_R^2}{s}\right)
+\frac{\beta_0}{6} \zeta_3 c_3-\frac{\beta_0}{4} \zeta_2 c_2
-\frac{\beta_0}{2} \zeta_2 C_F \ln\left(\frac{\mu_F^2}{\mu_R^2}\right)
\nonumber \\ &&
{}+2 A_q^{(2)} \zeta_2-2 \gamma_q^{(2)} \ln\left(\frac{\mu_F^2}{s}\right)
+\frac{\beta_0}{8} \left[ 2 \gamma_q^{(1)}-C_F \ln\left(\frac{t_1 u_1}{m^4}\right)
\right] \ln^2\left(\frac{\mu_F^2}{s}\right)
\nonumber \\ &&
{}+A_q^{(2)} \ln\left(\frac{t_1 u_1}{m^4}\right) \ln\left(\frac{\mu_F^2}{s}\right)
+\frac{\beta_0}{6} C_F \ln^3\left(\frac{-t_1}{m^2}\right)
+\frac{\beta_0}{6} C_F \ln^3\left(\frac{-u_1}{m^2}\right)
\nonumber \\ &&
{}+\left(C_F \frac{\beta_0}{4}+A_q^{(2)}\right)
\left[\ln^2\left(\frac{-t_1}{m^2}\right)
+\ln^2\left(\frac{-u_1}{m^2}\right)\right]\, ,
\label{X0}
\eeqa
and
\beqa
Y_0&=&-C_F \frac{\beta_0^2}{24} \left[\ln^3\left(\frac{\mu_F^2}{\mu_R^2}\right)+\ln^3\left(\frac{\mu_R^2}{s}\right)\right]+\frac{1}{16}\left(C_F \beta_1+8\beta_0 A_q^{(2)}\right) \left[\ln^2\left(\frac{\mu_F^2}{\mu_R^2}\right)-\ln^2\left(\frac{\mu_R^2}{s}\right)\right]
\nonumber \\ &&
{}-\left(A_q^{(2)} \beta_0+C_F \frac{\beta_1}{8}\right) \ln\left(\frac{t_1 u_1}{m^4}\right) \ln\left(\frac{\mu_R^2}{s}\right)+D_q^{(2)} \beta_0 \ln\left(\frac{\mu_R^2}{s}\right)-2 A_q^{(3)} \ln\left(\frac{\mu_F^2}{s}\right)+2 D_q^{(3)}
\nonumber \\ &&            
{}-\frac{C_F}{6} \beta_0^2 \left[\ln^3\left(\frac{-t_1}{m^2}\right)+\ln^3\left(\frac{-u_1}{m^2}\right)\right] -\frac{\beta_0}{4}\left(C_F \beta_0+4 A_q^{(2)}\right) \left[\ln^2\left(\frac{-t_1}{m^2}\right)+\ln^2\left(\frac{-u_1}{m^2}\right)\right] 
\nonumber \\ &&            
{}+\left(\beta_0 D_q^{(2)}-2A_q^{(3)}\right) \ln\left(\frac{t_1 u_1}{m^4}\right)
\label{Y0}
\eeqa
with $\beta_1=34 C_A^2/3-2 C_F n_f-10 C_A n_f/3$ \cite{WEC,DRTJ,ET79} the coefficient of the second term in the perturbative expansion of the $\beta$ function.
For processes $gg \to AB$, we replace in Eqs. (\ref{N3LO}), (\ref{X1}), (\ref{X0}), and (\ref{Y0}) all explicit appearances of $C_F$ by $C_A$, and we use the corresponding expressions for $A_g$, $D_g$, $\gamma_g$, $c_3$, $c_2$, and $c_1$ as detailed above, as well as the appropriate expressions for $V_1$ and $V_2$.

Finally, we consider the more general case where the observed particle $A$ has mass $m_A$, which is different from the mass $m_B$ of the other final-state particle $B$. In that case, we simply replace $m$ by $m_A$ everywhere it appears explicitly in Eqs. (\ref{Dplus}) through (\ref{Y0}), and we also redefine and use $t_1=t-m_B^2$ and $u_1=u-m_B^2$ everywhere. With these simple changes all the expressions above can be used for the more general case of unequal masses. 

It is worth emphasizing that the expressions for the higher-order soft-gluon corrections presented in this section are universal and can be applied to any electroweak production process that occurs through quark-antiquark annihilation or gluon-gluon fusion. The only process-dependent parts are encoded into the virtual correction terms $V_1$ and $V_2$, which depend on the specific process. For example, explicit expressions of those terms for various electroweak processes can be found in~\cite{NKhiggs}.

\mysection{Charged Higgs pair production in the 2HDM}

In this section we discuss the production of a charged Higgs pair via quark-antiquark annihilation in the 2HDM that occurs through diagrams of Drell-Yan type. For concreteness, we consider a CP-conserving type II 2HDM with a softly-broken $Z_2$ symmetry defined by the transformations $\Phi_1\to \Phi_1$ and  $\Phi_2\to -\Phi_2$, where $\Phi_1$ and $\Phi_2$ are the two Higgs doublets. In this model, the Higgs doublet $\Phi_1$ is responsible for the Yukawa interactions that generate masses for the down-type quarks  and charged leptons, while $\Phi_2$ is responsible for the Yukawa interactions that generate masses for the up-type quarks. The Higgs potential is given by:
\beqa\label{scalpot}
V_{\rm 2HDM}&=&m_1^2|\Phi_1|^2+m_2^2|\Phi_2|^2-m_{12}^2(\Phi_1^\dagger \Phi_2+\Phi_2^\dagger \Phi_1)+\frac{\lambda_1}{2}|\Phi_1|^4+\frac{\lambda_2}{2}|\Phi_2|^4\nonumber\\
&&+\lambda_3|\Phi_1|^2|\Phi_2|^2+\lambda_4|\Phi_1^\dagger\Phi_2|^2+\frac{\lambda_5}{2}[(\Phi_1^\dagger \Phi_2)^2+(\Phi_2^\dagger \Phi_1)^2]\,,
\eeqa
where all the masses ($m_1$, $m_2$, $m_{12}$) and quartic couplings ($\lambda_1$, $\lambda_2$, $\lambda_3$, $\lambda_4$, $\lambda_5$) are taken to be real in order to avoid CP violation. After electroweak symmetry breaking, this model predicts the existence of three neutral scalar fields (two CP-even states $h$ and $H$, and a CP-odd state $A$) and a pair of charged Higgs bosons $H^{\pm}$. The eight parameters of the scalar Higgs potential in Eq. (\ref{scalpot}) are usually rewritten in terms of more physical ones: a common choice is to consider the electroweak vacuum expectation value $\upsilon=246$ GeV, the four Higgs masses $m_h$, $m_H$, $m_A$, $m_{H^+}$, the tangent $\tan\beta$, the cosine $\cos(\beta-\alpha)$, and the discrete symmetry soft-breaking term $m_{12}$ (for more details see Section III of ~\cite{Kanemura:2004mg}). For this particular model, recent studies~\cite{Chowdhury:2017aav,Atkinson:2021eox,Wang:2022yhm,Arco:2022jrt} showed that the size of $\cos(\beta-\alpha)$ has to be lower than ${\cal O}(10^{-2})$ and that low values of $\tan\beta$ are favored. 

\begin{figure}[htbp]
\begin{center}
\includegraphics[width=180mm]{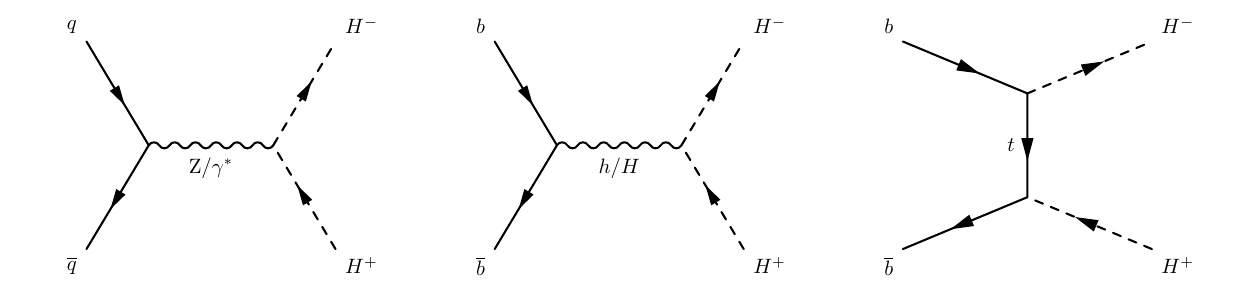}
\caption{LO Feynman diagrams for quark-induced $H^+ H^-$ production at the LHC.}
\label{dy_feynman}
\end{center}
\end{figure}

As already mentioned, charged-Higgs pair production is a process that can occur through quark-antiquark annihilation and gluon-fusion channels. In this work we consider only quark-induced production which is the dominant  production mechanism at low $\tan\beta$ values~\cite{Alves:2005kr}. Leading-order Feynman diagrams for quark-induced production are shown in Fig.~\ref{dy_feynman}. In the case of light quarks in the initial state, we have diagrams of Drell-Yan type where the production of a pair of charged Higgs bosons occurs through the exchange of a photon and a $Z$ boson in the $s$-channel (left diagram of Fig.~\ref{dy_feynman}). In the case of bottom quarks, in addition to Drell-Yan type diagrams, the production of charged Higgs bosons also occurs through the exchange of the CP-even neutral Higgs bosons ($h,H$) in the $s$-channel (middle diagram of Fig. \ref{dy_feynman}) and the top-quark in the $t$-channel (right diagram of Fig.~\ref{dy_feynman}). These extra diagrams might interfere with Drell-Yan type ones and introduce some dependence on the scalar potential parameters $\tan\beta$, $\cos(\beta-\alpha)$ and $m_{12}^2$.

In this work, we consider only Drell-Yan type production of a pair of charged Higgs bosons in the 5-flavor scheme where we include also a massless bottom quark in the initial state. In this way, our $H^+H^-$ production process depends only on the mass of the final-state charged Higgs and is independent of the other 2HDM scalar potential parameters, like $\tan\beta$, $\cos(\beta-\alpha)$, and $m_{12}^2$. In considering the total contribution of both light and heavy quarks in the initial state, we are neglecting some bottom-quark contributions (middle and right diagrams of Fig.~\ref{dy_feynman}), which might give a  sizable contribution, depending on the specific values of the Higgs potential parameters. However, for small values of $\tan \beta$ favored by recent fits, i.e. $2<\tan\beta<4$, these contributions are of the order of 2-3\% at LO and NLO, as also shown in~\cite{Alves:2005kr}, which are smaller than the aNNLO corrections for the Drell-Yan type processes and within the theoretical uncertainty of our calculation obtained from scale and pdf variation. We would like to emphasize that the goal of this section is to provide the most precise prediction for processes that occur through Drell-Yan type diagrams, which for light quarks represent their complete contribution. Providing the total $b\bar b$ contribution, which depends on the values of several Higgs potential parameters, is beyond the scope of this paper. 

Therefore, the quantity $F^B_{q{\bar q} \to AB}$ entering in the Born differential cross section in Eq.~(\ref{LO}) computed using Drell-Yan type diagrams is given by 
\beqa
F^B_{u{\bar u} \to H^+ H^-}&=&\frac{1}{16\pi s^2}\frac{\alpha^2 \pi^2 m_Z^4}{27 m_W^4}\frac{t_1 \, u_1+m_{H^+}^2(t_1+u_1)}{(m_W^2-m_Z^2)^2(m_Z^2-s)^2s^2} \nonumber\\ && \times \left[128 m_W^8+17m_Z^4s^2-32m_W^6(8m_Z^2+s)-4m_W^2m_Z^2s(20m_Z^2+7s) \right.
\nonumber\\ && \left. \quad \; {}+4m_W^4(32m_Z^4+28m_Z^2s+5s^2)\right]
\eeqa 
for up-type quarks, while for down-type quarks we have
\beqa
F^B_{d{\bar d} \to H^+ H^-}&=&\frac{1}{16\pi s^2}\frac{\alpha^2 \pi^2 m_Z^4}{27 m_W^4}\frac{t_1 \, u_1+m_{H^+}^2(t_1+u_1)}{(m_W^2-m_Z^2)^2(m_Z^2-s)^2s^2} 
\nonumber\\ && \times \left[32 m_W^8+5m_Z^4s^2+16m_W^6(-4m_Z^2+s)-8m_W^2m_Z^2s(m_Z^2+2s) \right.
\nonumber\\ && \left. \quad \; 
{}+4m_W^4(8m_Z^4-2m_Z^2s+5s^2)\right]\,.
\eeqa 
These expressions are used in Eq. (\ref{LO}) to compute the Born cross section, and in Eqs. (\ref{NLO}), (\ref{NNLO}), and (\ref{N3LO}) to derive the higher-order soft-gluon corrections through N$^3$LO.

\mysection{Total cross sections for $H^+H^-$ production at the LHC}

In this section we present results for the total cross section of charged-Higgs pair production via quark-antiquark annihilation $q\bar q \to H^+ H^-$ at the LHC, for two center-of-mass energies, namely 13 TeV and 13.6 TeV. As already discussed in the previous section, this cross section depends only on $m_{H^+}$ and is independent of the other Higgs potential parameters. In our calculation we work in the 5-flavor scheme and we include the bottom quark in the initial state, which is taken to be massless. Lower bounds on the mass of the charged Higgs of the order of 570-800 GeV have been obtained in Ref.~\cite{Misiak:2017bgg}. In light of these constraints, the charged-Higgs mass is taken here to vary between 500 and 1500 GeV. Lower values of $m_{H^+}$ are excluded, while higher values, although allowed, are not considered because they give very small cross sections. The complete LO and NLO QCD results are calculated using {\small \sc MadGraph5\_aMC@NLO} \cite{MG5} and an ad hoc UFO model for the CP-conserving type II 2HDM, which can be found at~\cite{WuModel}. We take $m_W=80.377$ GeV, $m_Z=91.1876$ GeV, and $\alpha^{-1}=127.9$. We use the same MSHT20 aN$^3$LO pdf set \cite{MSHT20} for computing results at every perturbative order, in order to show how each order in the series contributes to the aN$^3$LO cross section.  

For the total cross section, the central results are obtained by setting the factorization and renormalization scales both equal to $m_{H^+}$. Scale uncertainties are obtained by varying independently $\mu_R$  and  $\mu_F$ between $m_{H^+}/2$ and  $2 m_{H^+}$, and pdf uncertainties are also computed. The cross sections at aNNLO are computed by adding the second-order soft-plus-virtual QCD corrections to the complete NLO result. The third-order soft-gluon corrections are further added to derive the result at aN$^3$LO.

\begin{figure}[htb!]
\begin{center}
\includegraphics[width=120mm]{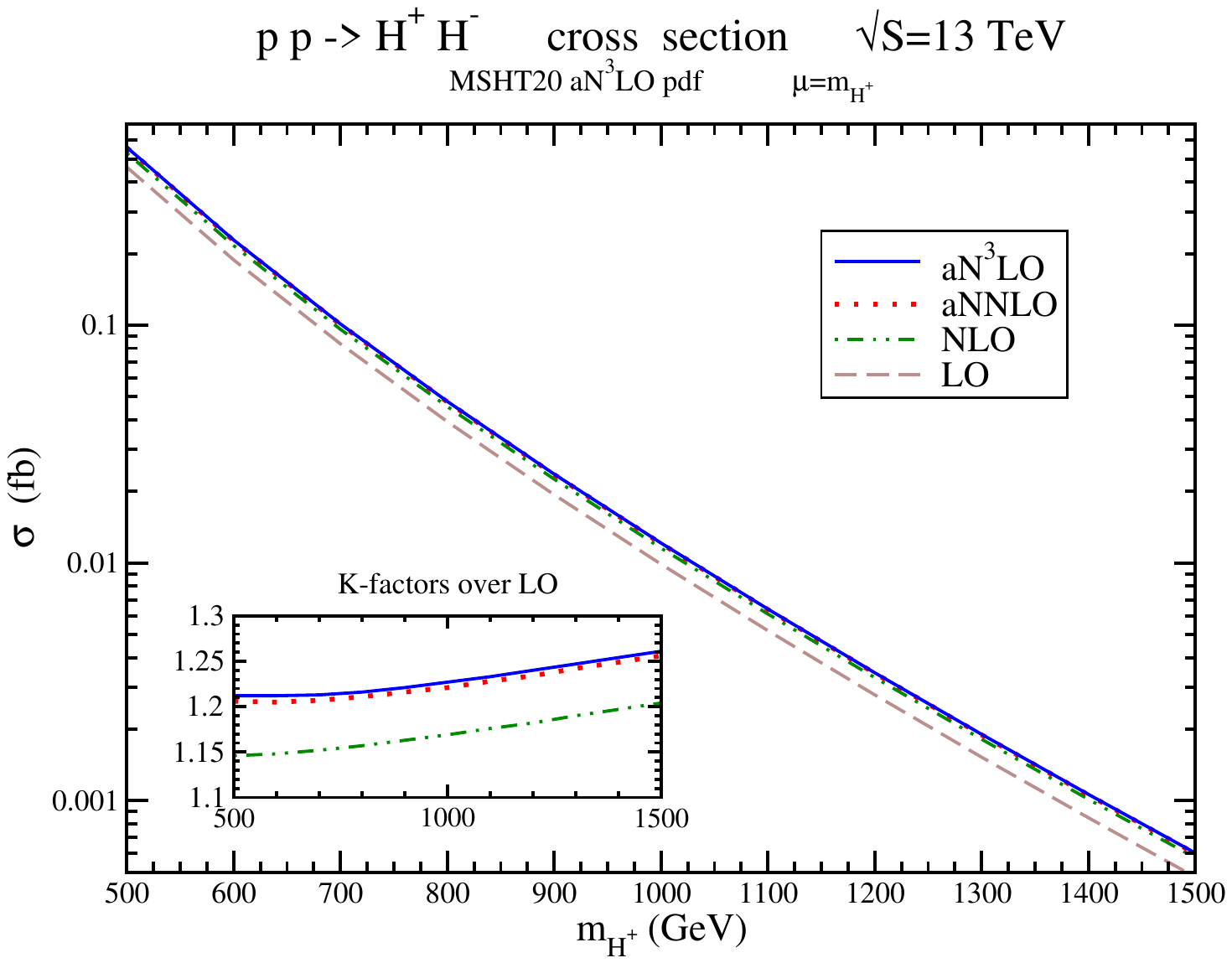}
\caption{The total cross sections at LO, NLO, aNNLO, and aN$^3$LO for $q\bar q\to H^+ H^-$ production in $pp$ collisions at 13 TeV energy with $\mu_F=\mu_R=m_{H^+}$ using MSHT20 aN$^3$LO pdf. The inset plot displays the NLO/LO, aNNLO/LO, and aN$^3$LO/LO $K$-factors.}
\label{xsec13tev}
\end{center}
\end{figure}

\begin{figure}[htb!]
\begin{center}
\includegraphics[width=120mm]{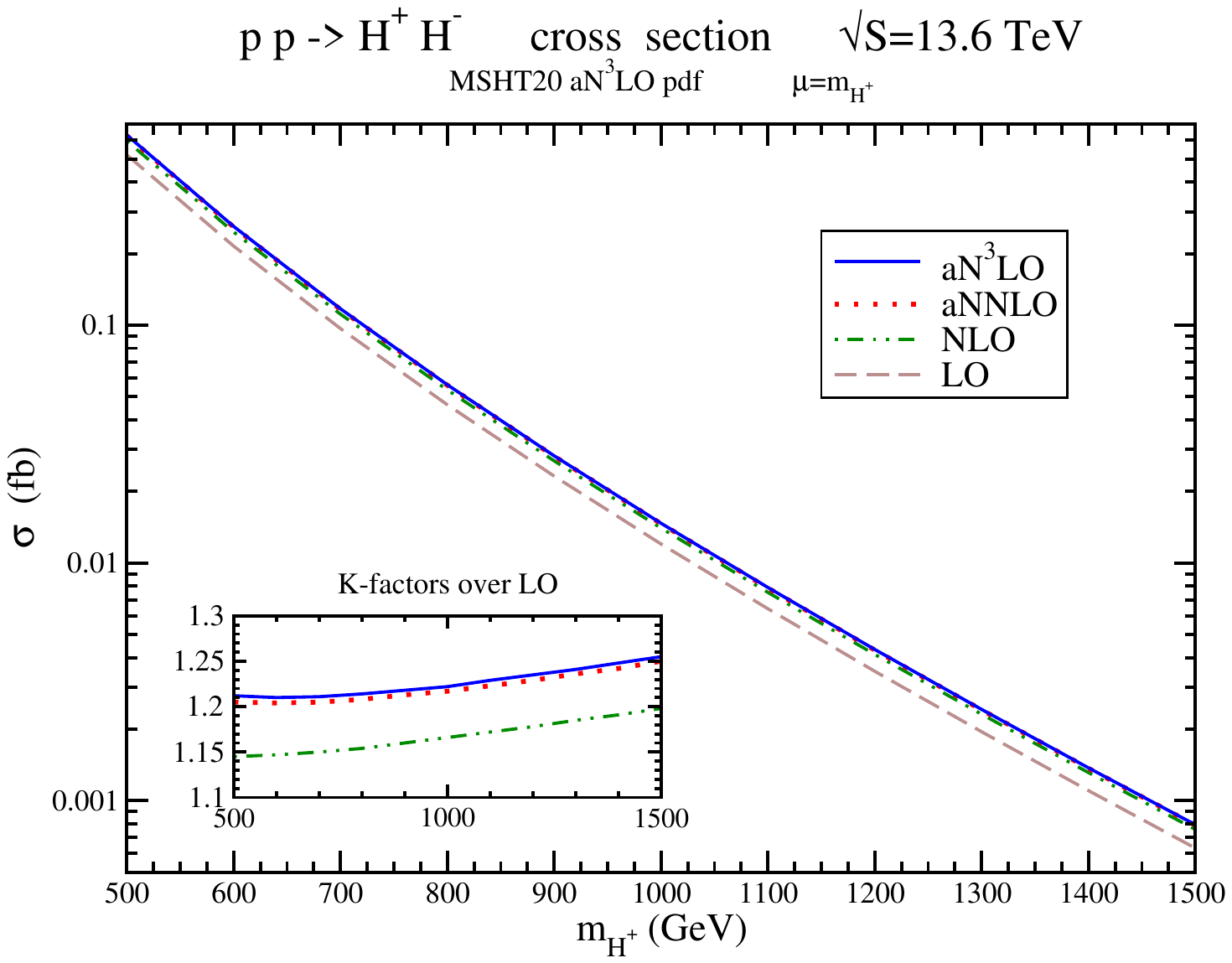}
\caption{The total cross sections at LO, NLO, aNNLO, and aN$^3$LO for $q\bar q\to H^+ H^-$ production in $pp$ collisions at 13.6 TeV energy with $\mu_F=\mu_R=m_{H^+}$ using MSHT20 aN$^3$LO pdf. The inset plot displays the NLO/LO, aNNLO/LO, and aN$^3$LO/LO $K$-factors.}
\label{xsec136tev}
\end{center}
\end{figure}

In Fig. 2 and Fig. 3 we show the total cross section for $ H^+ H^-$  production at the LHC with 13 TeV and 13.6 TeV energy, respectively. These plots show the central value, obtained with $\mu=m_{H^+}$ (where $\mu=\mu_R=\mu_F$), of the LO, NLO, aNNLO, and aN$^3$LO cross sections as a function of the mass of the charged Higgs boson. The inset plots display the $K$-factors of the higher-order cross sections relative to the LO result. At each perturbative order, we find that the $K$-factors slowly increase with increasing charged-Higgs mass. For instance, at both 13 TeV and 13.6 TeV, the NLO $K$-factor varies from 1.15 for $m_{H^+}=500$ GeV to 1.20 for $m_{H^+}=1500$ GeV. The NNLO soft-plus-virtual corrections provide an additional 5\% to 6\% increase, and the N$^3$LO soft-gluon corrections give a further 0.5\% to 0.6\% increase. Thus, the aN$^3$LO $K$-factor varies from 1.21 for $m_{H^+}=500$ GeV to 1.26 for $m_{H^+}=1500$ GeV.

\begin{table}[htb!]
\begin{center}
\begin{tabular}{|c|c|c|c|c|} \hline
\multicolumn{5}{|c|}{$p p \to H^+ H^-$ cross section via quark-antiquark annihilation at 13 TeV} \\ \hline
 \!\!$m_{H^+} \, {\rm (GeV)}$\!\! &  $\sigma$ LO (fb) & $\sigma$ NLO (fb)  & $\sigma$ aNNLO (fb) & $\sigma$ aN$^3$LO (fb)\\ \hline
$600$ & $0.188^{+0.016}_{-0.014}{}^{+0.006}_{-0.006}$ & $0.216^{+0.005}_{-0.006}{}^{+0.007}_{-0.007}$ & $0.227^{+0.003}_{-0.001}{}^{+0.007}_{-0.007}$ & $0.228^{+0.003}_{-0.002}{}^{+0.007}_{-0.007}$ \\ \hline
$800$ & $0.0393^{+0.0040}_{-0.0034}{}^{+0.0017}_{-0.0015}$ & $0.0454^{+0.0011}_{-0.0013}{}^{+0.0019}_{-0.0018}$ & $0.0476^{+0.0004}_{-0.0002}{}^{+0.0020}_{-0.0019}$ & $0.0478^{+0.0006}_{-0.0002}{}^{+0.0020}_{-0.0019}$ \\ \hline
$1000$ & $0.00988^{+0.00114}_{-0.00096}{}^{+0.00052}_{-0.00049}$ & $0.01155^{+0.00030}_{-0.00038}{}^{+0.00060}_{-0.00057}$ & $0.01207^{+0.00008}_{-0.00010}{}^{+0.00063}_{-0.00060}$ &$0.01212^{+0.00009}_{-0.00003}{}^{+0.00063}_{-0.00060}$ \\ \hline
$1200$ & $0.00279^{+0.00035}_{-0.00030}{}^{+0.00019}_{-0.00018}$ & $0.00329^{+0.00009}_{-0.00012}{}^{+0.00022}_{-0.00021}$ & $0.00344^{+0.00002}_{-0.00004}{}^{+0.00023}_{-0.00022}$ &$0.00345^{+0.00001}_{-0.00001}{}^{+0.00023}_{-0.00022}$ \\ \hline
\end{tabular}
\caption[]{The total cross section for $p p \to H^+ H^-$ via quark-antiquark annihilation at LO, NLO, aNNLO, and aN$^3$LO at the LHC with $\sqrt{S}=13$ TeV and MSHT20 aN$^3$LO pdf. The central results are for $\mu_F=\mu_R=m_{H^+}$ and are shown together with scale and pdf uncertainties.}
\label{table13TeVan3lopdf}
\end{center}
\end{table}

\begin{table}[htb!]
\begin{center}
\begin{tabular}{|c|c|c|c|c|} \hline
\multicolumn{5}{|c|}{$p p \to H^+ H^-$ cross section via quark-antiquark annihilation at 13.6 TeV} \\ \hline
\!\!$m_{H^+} \, {\rm (GeV)}$\!\! &  $\sigma$ LO (fb) & $\sigma$ NLO (fb)  & $\sigma$ aNNLO (fb) & $\sigma$ aN$^3$LO (fb) \\ \hline
$600$ & $0.215^{+0.017}_{-0.015}{}^{+0.007}_{-0.007}$ & $0.247^{+0.006}_{-0.006}{}^{+0.008}_{-0.007}$ & $0.259^{+0.003}_{-0.002}{}^{+0.008}_{-0.008} $ & $0.260^{+0.004}_{-0.002}{}^{+0.008}_{-0.008} $\\ \hline
$800$ & $0.0463^{+0.0046}_{-0.0039}{}^{+0.0018}_{-0.0017}$ & $0.0534^{+0.0012}_{-0.0016}{}^{+0.0021}_{-0.0020}$ & $0.0559^{+0.0005}_{-0.0003}{}^{+0.0022}_{-0.0021} $ & $0.0562^{+0.0007}_{-0.0003}{}^{+0.0022}_{-0.0021} $ \\ \hline
$1000$ & $0.0120^{+0.0014}_{-0.0011}{}^{+0.0006}_{-0.0006}$ & $0.0140^{+0.0004}_{-0.0005}{}^{+0.0007}_{-0.0007}$ & $0.0146^{+0.0001}_{-0.0001}{}^{+0.0007}_{-0.0007} $ & $0.0147^{+0.0001}_{-0.0001}{}^{+0.0007}_{-0.0007} $ \\ \hline
$1200$ & $0.00350^{+0.00043}_{-0.00036}{}^{+0.00023}_{-0.00022}$ & $0.00413^{+0.00011}_{-0.00015}{}^{+0.00026}_{-0.00025}$ & $0.00431^{+0.00003}_{-0.00004}{}^{+0.00027}_{-0.00026} $ & $0.00432^{+0.00002}_{-0.00001}{}^{+0.00027}_{-0.00026} $ \\ \hline
\end{tabular}
\caption[]{The total cross section for $p p \to H^+ H^-$ via quark-antiquark annihilation at LO, NLO, aNNLO, and aN$^3$LO at the LHC with $\sqrt{S}=13.6$ TeV and MSHT20 aN$^3$LO pdf. The central results are for $\mu_F=\mu_R=m_{H^+}$ and are shown together with scale and pdf uncertainties.}
\label{table136TeVan3lopdf}
\end{center}
\end{table}

In Table 1 and Table 2 we present some numerical values for the total cross section of $ H^+ H^-$ production in proton-proton collisions, with a center-of-mass energy of 13 TeV and 13.6 TeV, respectively. These cross sections are shown for select charged-Higgs masses, namely 600, 800, 1000, and 1200 GeV. They are calculated at LO, NLO, aNNLO, and aN$^3$LO and they are shown together with scale and pdf uncertainties.  We want to stress that the soft-gluon corrections are numerically dominant. The result derived with only NLO soft-gluon corrections differs from the exact NLO cross section by only around one percent or less for the masses and energies considered here. Furthermore, the aNNLO result would change by only 2 per mille if we did not include the $\delta(s_4)$ terms (i.e. the virtual corrections) at NNLO; the soft-gluon corrections are numerically by far the dominant contribution to the higher-order QCD corrections.   

As shown in Table 1 for the results at 13 TeV energy, the scale uncertainty at LO is large and is reduced significantly at NLO. The higher-order soft-plus-virtual corrections at NNLO decrease the uncertainty even more. Finally, the addition of the N$^3$LO soft-gluon corrections further decreases this uncertainty for larger charged-Higgs masses but not for smaller ones. The pdf uncertainty varies from about $\pm 3\%$ for a mass of 600 GeV to $\pm 7\%$ for a 1200 GeV mass, at each perturbative order. The results in Table 2 at 13.6 TeV energy show similar behavior for the scale uncertainty at different perturbative orders. Also in this case, the pdf uncertainty varies at each perturbative order from about $\pm 3\%$  for $m_{H^+}=600$ GeV to $\pm 6\%$ for $m_{H^+}=1200$ GeV.

Regarding pdf uncertainties, we can see that, for both center-of-mass energies, they increase with increasing mass of the charged Higgs boson. Moreover, they become bigger than the scale ones already at NLO and therefore, starting at that order, they should be considered the dominant source of theoretical uncertainty. As argued in~\cite{MSHT20}, with the use of aN$^3$LO pdf, the factorisation scale variation is already contained within the predicted pdf uncertainties.

It is also interesting to study the relative contributions of the different logarithmic powers to the higher-order corrections. As an example, we consider a charged-Higgs mass of 600 GeV and an energy of 13.6 TeV. Setting $\mu_F=\mu_R=m_{H^+}$ and using the notation of Eq. (\ref{Dplus}), the NNLO corrections from the ${\cal D}_3$ terms alone are 0.0134 fb, from the ${\cal D}_3$ terms plus the ${\cal D}_2$ terms they are 0.0189 fb, from ${\cal D}_3+{\cal D}_2+{\cal D}_1$ they are 0.0150 fb, and from ${\cal D}_3+{\cal D}_2+{\cal D}_1+{\cal D}_0$ they are 0.0118 fb. By further adding the NNLO $\delta(s_4)$ terms, we get the final NNLO soft-plus-virtual correction of 0.0123 fb. Thus, we see that the virtual contribution to the NNLO soft-plus-virtual corrections is small, and the soft-gluon corrections alone are dominant. We also see that the leading-logarithmic contribution (i.e. from ${\cal D}_3$) alone is quite close to the complete soft-plus-virtual contribution at NNLO.

For the N$^3$LO soft-gluon contributions, again with a 600 GeV charged-Higgs mass at 13.6 TeV energy, we find a contribution of 0.00993 fb from the ${\cal D}_5$ terms alone, 0.0176 fb from ${\cal D}_5+{\cal D}_4$, 0.0155 fb from ${\cal D}_5+{\cal D}_4+{\cal D}_3$, 0.00898 fb from ${\cal D}_5+{\cal D}_4+{\cal D}_3+{\cal D}_2$, 0.00200 from 
${\cal D}_5+{\cal D}_4+{\cal D}_3+{\cal D}_2+{\cal D}_1$, and finally 0.00133 fb from ${\cal D}_5+{\cal D}_4+{\cal D}_3+{\cal D}_2+{\cal D}_1+{\cal D}_0$ which is the total N$^3$LO soft-gluon correction. Thus, at N$^3$LO, the leading-logarithmic contribution (i.e. from ${\cal D}_5$) is very different from the complete soft-gluon correction, in contrast to what we found above for the NNLO corrections. We also see that the N$^3$LO soft-gluon corrections are an order of magnitude smaller than those at NNLO.

As mentioned in the previous section, for small values of $\tan \beta$, the additional contributions from $b{\bar b} \to H^+ H^-$ are rather small and less than the aNNLO corrections of the Drell-Yan type $q{\bar q} \to H^+ H^-$ diagrams. One can calculate these $b{\bar b}$ contributions at NLO for any choice of the Higgs-potential parameters, and then add them to our aN$^3$LO result for the Drell-Yan type $q{\bar q}$ contributions. For example, for $\tan \beta=3$, $m_{H^+}=600$ GeV, $m_H=400$ GeV, $\cos(\beta-\alpha)=0$, and $m_{12}=0$ GeV, the missing $b{\bar b}$ contribution at 13 TeV energy is 0.0035 fb. Looking at the first line of Table 1, we see that this is much smaller than the aNNLO corrections of 0.011 fb for $q{\bar q}$ as well as the total uncertainty of the overall $q{\bar q}$ cross section. Adding the $b{\bar b}$ contribution to the aN$^3$LO $q{\bar q}$ cross section gives a total of 0.232 fb. Clearly, in this case it is not necessary to calculate the aNNLO and aN$^3$LO corrections for the $b{\bar b}$ contribution since they would be much smaller than the precision of the numbers displayed.  

\mysection{Charged-Higgs rapidity distributions}

As the formalism presented in Section 2 indicates, soft-gluon resummation is derived not only for total cross sections but also for differential distributions. In this section we present results for the charged-Higgs rapidity distribution in $H^+ H^-$ production at the LHC with 13.6 TeV center-of-mass energy, for two specific charged-Higgs masses, namely $m_{H^+}=600$ GeV and $m_{H^+}=800$ GeV. Again, we use {\small \sc MadGraph5\_aMC@NLO} \cite{MG5} for the complete NLO results to which we add the second-order soft-plus-virtual corrections to derive aNNLO distributions. The third-order soft-gluon corrections are also added in order to derive results at aN$^3$LO.

\begin{figure}[htb!]
\begin{center}
\includegraphics[width=120mm]{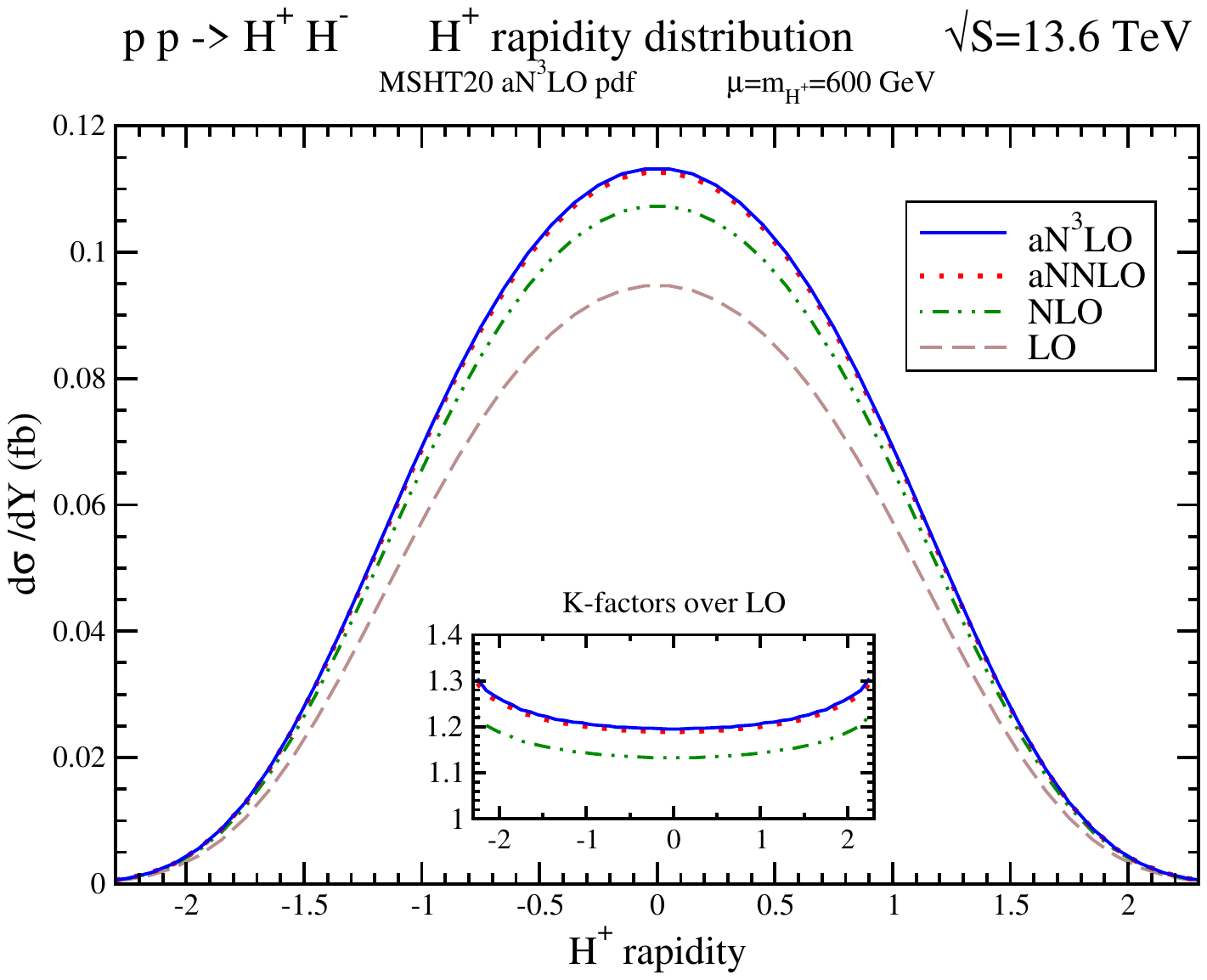}
\caption{The rapidity distribution of the $H^+$ with a mass of 600 GeV at LO, NLO, aNNLO, and aN$^3$LO for $q\bar q\to H^+ H^-$ production in $pp$ collisions at 13.6 TeV energy with $\mu_F=\mu_R=m_{H^+}$ using MSHT20 aN$^3$LO pdf. The inset plot displays the NLO/LO, aNNLO/LO, and aN$^3$LO/LO $K$-factors.}
\label{rapidity600}
\end{center}
\end{figure}

\begin{figure}[htb!]
\begin{center}
\includegraphics[width=120mm]{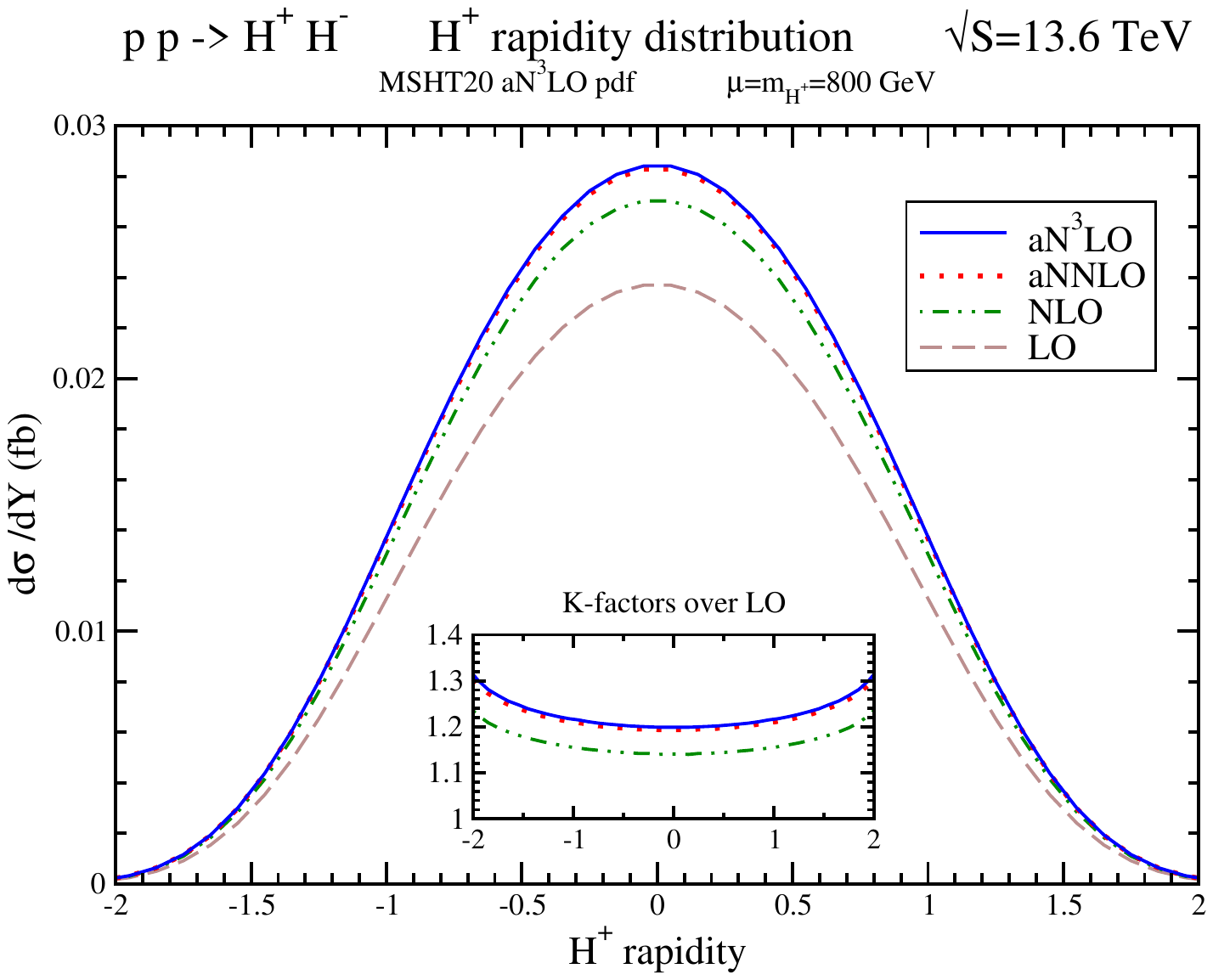}
\caption{The rapidity distribution of the $H^+$ with a mass of 800 GeV at LO, NLO, aNNLO, and aN$^3$LO for $q\bar q\to H^+ H^-$ production in $pp$ collisions at 13.6 TeV energy with $\mu_F=\mu_R=m_{H^+}$ using MSHT20 aN$^3$LO pdf. The inset plot displays the NLO/LO, aNNLO/LO, and aN$^3$LO/LO $K$-factors.}
\label{rapidity800}
\end{center}
\end{figure}
 
In Fig. 4, we plot the central ($\mu=m_{H^+}$) results at LO, NLO, aNNLO, and aN$^3$LO for the $H^+$ rapidity distribution in $H^+ H^-$ production at 13.6 TeV LHC energy with MSHT20 aN$^3$LO pdf, for $m_{H^+}$ equal to $600$ GeV.
The $K$-factors relative to the LO results are shown in the inset plot. The enhancement from the aNNLO soft-plus-virtual corrections is  significant while the aN$^3$LO soft-gluon contribution is smaller than 1\% in the rapidity range shown in the figure, similar to the total cross section case. We see that the $K$-factors are relatively flat for central values of the charged-Higgs rapidity, but they begin to grow for rapidities larger than 1. We also note that the distribution becomes very small for rapidity values above 2.

Regarding the theoretical uncertainties from scale variation, we note that they are essentially the same as those for the total cross section in the rapidity range plotted, and only become bigger at very large rapidities, where the numerical value of the distribution is very small. As for the total cross section, the higher-order corrections reduce the scale dependence. The pdf uncertainties are also similar to those of the total cross section, around 3\%, in the rapidity range shown in the plot, but they increase at larger rapidities. 

In Fig. 5, we plot the central ($\mu=m_{H^+}$) results at LO, NLO, aNNLO, and aN$^3$LO for the $H^+$ rapidity distribution in $H^+ H^-$ production at 13.6 TeV LHC energy with MSHT20 aN$^3$LO pdf, for $m_{H^+}=800$ GeV. Although the numerical values for the distribution are much smaller than those found with the  smaller mass in Fig. 4, the $K$-factors relative to the LO results in the inset plot show a similar pattern, i.e. they are relatively flat for central values of the charged-Higgs rapidity but begin to grow for rapidities larger than 1. Also, the theoretical uncertainties from scale variation are again essentially the same as those for the total cross section in the rapidity range plotted, and only become bigger at very large rapidities, where the distribution is negligible. Finally, the pdf uncertainties are also similar to those of the total cross section for an 800 GeV mass, around 4\%, in the rapidity range shown in the plot, but they increase at larger rapidities.

\mysection{Conclusions}

In this paper we have presented the first calculation of the complete soft-gluon corrections in the production of colorless final states through N$^3$LO in single-particle-inclusive kinematics. We have derived the detailed analytical expressions for these corrections and, as a concrete application, we have used our results to study higher-order QCD corrections for the production of a charged Higgs pair ($H^+ H^-$ production) at the LHC via quark-antiquark annihilation in the 2HDM. This calculation is particularly relevant for type II 2HDM, where the value of $\tan\beta$ is constrained by experiments to be small and the quark-antiquark  annihilation production mechanism is indeed the dominant channel  for such small values of $\tan\beta$. 

We have computed cross sections at aNNLO by adding the second-order soft-plus-virtual QCD corrections to the complete NLO result, and at aN$^3$LO by further adding the third-order soft-gluon corrections. 
Our results show that the NLO corrections increase the LO cross section by 15\% to 20\% for charged-Higgs masses in the range from 500 to 1500 GeV, with the exact numbers depending on the collision energy and the charged-Higgs mass, while the NNLO soft-plus-virtual corrections provide an additional 5\% to 6\% increase. The N$^3$LO soft-gluon corrections provide a further increase of 0.5\% to 0.6\%.
In general, the soft-gluon corrections are by far the dominant contribution to the higher-order QCD corrections. For instance, the exact NLO cross section differs by only around one percent or less from the one with NLO soft-gluon corrections alone, and the aNNLO result would change by only 2 per mille if we did not include the virtual corrections at NNLO. For the total cross section, we find that the theoretical uncertainties from scale variation get substantially reduced by going to higher orders.

We have also computed aNNLO and aN$^3$LO charged-Higgs rapidity distributions at the LHC with 13.6 TeV center-of-mass energy, for two representative masses, namely $m_{H^+}=600$ GeV and $m_{H^+}=800$ GeV. We found that the enhancement from the aNNLO soft-plus-virtual corrections is  significant while the aN$^3$LO soft-gluon contribution is smaller, similar to the total cross section case. The $K$-factors are relatively flat for central values of the charged-Higgs rapidity, but begin to grow for rapidities larger than one. The theoretical uncertainties from scale variation and from the pdf are basically the same as those of the total cross section for rapidities less than two, where the rapidity distribution is not negligible.

Finally, we want to emphasize that this work opens the way for using soft-gluon corrections through N$^3$LO in single-particle-inclusive kinematics to calculate more precise predictions of production rates of colorless final states in hadron-hadron collisions, both in the Standard Model and beyond.

\section*{Acknowledgements}
This material is based upon work supported by the National Science Foundation under Grant No. PHY 2112025.


\begin{thebibliography}{99}

\bibitem{GS87}
G. Sterman, {\sl Summation of large corrections to short-distance hadronic cross sections}, Nucl. Phys. B {\bf 281}, 310 (1987).

\bibitem{CT89}
S. Catani and L. Trentadue, {\sl Resummation of the QCD perturbative series for hard processes}, Nucl. Phys. B {\bf 327}, 323 (1989).

\bibitem{NKGS1} 
N. Kidonakis and G. Sterman, {\sl Subleading logarithms in QCD hard scattering}, Phys. Lett. B {\bf 387}, 867 (1996). 

\bibitem{NKGS2} 
N. Kidonakis and G. Sterman, {\sl Resummation for QCD hard scattering}, Nucl. Phys. B {\bf 505}, 321 (1997) [arXiv:hep-ph/9705234].

\bibitem{LOS}
E. Laenen, G. Oderda, and G. Sterman,
{\sl Resummation of threshold corrections for single-particle inclusive cross sections},
Phys. Lett. B {\bf 438}, 173 (1998) [arXiv:hep-ph/9806467].

\bibitem{NK2loop} 
N. Kidonakis, {\sl Two-loop soft anomalous dimensions and next-to-next-to-leading-logarithm resummation for heavy quark production}, Phys. Rev. Lett. {\bf 102}, 232003 (2009) [arXiv:0903.2561].

\bibitem{NKnnll}
N. Kidonakis, {\sl Next-to-next-to-leading soft-gluon corrections for the top quark cross section and transverse momentum distribution}, Phys. Rev. D {\bf 82}, 114030 (2010) [arXiv:1009.4935].

\bibitem{NKan3lo}
N. Kidonakis, {\sl NNNLO soft-gluon corrections for the top-antitop pair production cross section}, Phys. Rev. D {\bf 90}, 014006 (2014) [arXiv:1405.7046].

\bibitem{MA05}
S. Moch and A. Vogt, {\sl Higher-order soft corrections to lepton pair and Higgs boson production}, Phys. Lett. B {\bf 631}, 48 (2005) [arXiv:hep-ph/0508265].

\bibitem{VR06}
V. Ravindran, {\sl Higher-order threshold effects to inclusive processes in QCD}, Nucl. Phys. B {\bf 752}, 173 (2006) [arXiv:hep-ph/0603041].

\bibitem{NKhiggs}
N. Kidonakis, {\sl Collinear and soft-gluon corrections to Higgs production at next-to-next-to-next-to-leading order}, Phys. Rev. D {\bf 77}, 053008 (2008) [arXiv:0711.0142]. 

\bibitem{2HDMreview}
G.C.~Branco, P.M.~Ferreira, L.~Lavoura, M.N.~Rebelo, M.~Sher, and J.P.~Silva, {\sl Theory and phenomenology of two-Higgs-doublet models},
Phys. Rept. {\bf 516}, 1 (2012) [arXiv:1106.0034].

\bibitem{NKcH}
N. Kidonakis, {\sl Charged Higgs production via $bg \to tH^-$ at the LHC}, JHEP {\bf 05}, 011 (2005) [arXiv:hep-ph/0412422]. 

\bibitem{NKtWH}
N. Kidonakis, {\sl Two-loop soft anomalous dimensions for single top quark associated production with a $W^-$ or $H^-$}, Phys. Rev. D {\bf 82}, 054018 (2010) [arXiv:1005.4451]. 

\bibitem{NKcHnnlo}
N. Kidonakis, {\sl Charged Higgs production in association with a top quark at approximate NNLO}, Phys. Rev. D {\bf 94}, 014010 (2016) [arXiv:1605.00622]. 

\bibitem{NKWH}
N. Kidonakis, {\sl Higher-order radiative corrections for $b{\bar b} \to H^- W^+$}, Phys. Rev. D {\bf 97}, 034002 (2018) [arXiv:1704.08549]. 

\bibitem{Djouadi:1999ht}
A.~Djouadi and M.~Spira, {\sl Supersymmetric QCD corrections to Higgs boson production at hadron colliders},
Phys. Rev. D {\bf 62}, 014004 (2000) [arXiv:hep-ph/9912476].

\bibitem{Hou:2005alu}
H.S.~Hou, W.G.~Ma, R.Y.~Zhang, Y.~Jiang, L.~Han, and L.R.~Xing, {\sl Pair production of charged Higgs bosons from bottom-quark fusion}, Phys. Rev. D {\bf 71}, 075014 (2005) [arXiv:hep-ph/0502214].

\bibitem{Alves:2005kr}
A.~Alves and T.~Plehn, {\sl Charged Higgs boson pairs at the CERN LHC}, Phys. Rev. D {\bf 71}, 115014 (2005)
[arXiv:hep-ph/0503135].

\bibitem{Jiang:1997cg}
Y.~Jiang, L.~Han, W.G.~Ma, Z.H.~Yu, and M.~Han, {\sl Pair production of charged Higgs bosons in gluon-gluon collisions}, J. Phys. G {\bf 23}, 385 (1997) [(E) {\bf 23}, 1151 (1997)] [arXiv:hep-ph/9703275].

\bibitem{Krause:1997rc}
A.~Krause, T.~Plehn, M.~Spira, and P.M.~Zerwas, {\sl Production of charged Higgs-boson pairs in gluon-gluon collisions}, Nucl. Phys. B {\bf 519}, 85 (1998) [arXiv:hep-ph/9707430].

\bibitem{Jiang:1997cr}
Y.~Jiang, W.G.~Ma, L.~Han, M.~Han, and Z.H.~Yu, {\sl SUSY charged Higgs boson pair production via gluon-gluon collisions}, J. Phys. G {\bf 24}, 83 (1998) [arXiv:hep-ph/9708421].

\bibitem{BarrientosBendezu:1999gp} 
A.A.~Barrientos Bendezu and B.A.~Kniehl, {\sl $H^+ H^-$ pair production at the Large Hadron Collider},
Nucl. Phys. B {\bf 568}, 305 (2000) [arXiv:hep-ph/9908385].

\bibitem{Brein:1999sy}
O.~Brein and W.~Hollik, {\sl Pair production of charged MSSM Higgs bosons by gluon fusion},
Eur. Phys. J. C {\bf 13}, 175 (2000) [arXiv:hep-ph/9908529].

\bibitem{Chowdhury:2017aav}
D.~Chowdhury and O.~Eberhardt, {\sl Update of global Two-Higgs-Doublet model fits}, JHEP {\bf 05}, 161 (2018) [arXiv:1711.02095].

\bibitem{Atkinson:2021eox}
O.~Atkinson, M.~Black, A.~Lenz, A.~Rusov, and J.~Wynne, {\sl Cornering the Two Higgs Doublet Model Type II}, JHEP {\bf 04}, 172 (2022) [arXiv:2107.05650].

\bibitem{Wang:2022yhm}
L.~Wang, J.M.~Yang, and Y.~Zhang, {\sl Two-Higgs-doublet models in light of current experiments: a brief review},
Commun. Theor. Phys. {\bf 74}, 097202 (2022) [arXiv:2203.07244].

\bibitem{Arco:2022jrt}
F.~Arco, S.~Heinemeyer, and M.J.~Herrero, {\sl Sensitivity and constraints to the 2HDM soft-breaking $Z_2$ parameter $m_{12}$}, Phys. Lett. B \textbf{835}, 137548 (2022) [arXiv:2207.13501].

\bibitem{NKVD}
N. Kidonakis and V. Del Duca, {\sl Electroweak-boson hadroproduction at large transverse momentum: factorization, resummation, and NNLO corrections}, Phys. Lett. B {\bf 480}, 87 (2000) [arXiv:hep-ph/9911460]. 

\bibitem{KT82}
J. Kodaira and L. Trentadue, {\sl Summing soft emission in QCD}, Phys. Lett. B {\bf 112}, 66 (1982). 

\bibitem{MVV04}
S. Moch, J.A.M. Vermaseren, and A. Vogt, {\sl The three-loop splitting functions in QCD: the non-singlet case}, Nucl. Phys. B {\bf 688}, 101 (2004)
[arXiv:hep-ph/0403192].

\bibitem{CLS97}
H. Contopanagos, E. Laenen, and G. Sterman, {\sl Sudakov factorization and resummation}, Nucl. Phys. B {\bf 484}, 303 (1997) [arXiv:hep-ph/9604313].

\bibitem{FRS1}
E.G. Floratos, D.A. Ross, and C.T. Sachrajda, {\sl Higher-order effects in asymptotically free gauge theories: The anomalous dimensions of Wilson operators}, Nucl. Phys. B {\bf 129}, 66 (1977) [(E) {\bf 139}, 545 (1978)].

\bibitem{FRS2}
E.G. Floratos, D.A. Ross, and C.T. Sachrajda, {\sl Higher-order effects in asymptotically free gauge theories: (II). Flavour singlet Wilson operators and coefficient functions}, Nucl. Phys. B {\bf 152}, 493 (1979).

\bibitem{GALY79}
A. Gonzalez-Arroyo, C. Lopez, and F.J. Yndurain,
{\sl Second-order contributions to the structure functions in deep inelastic scattering (I). Theoretical calculations}, Nucl. Phys. B {\bf 153}, 161 (1979).
 
\bibitem{GFP80}
G. Curci, W. Furmanski, and R. Petronzio, {\sl Evolution of parton densities beyond leading order: The non-singlet case}, Nucl. Phys. B {\bf 175}, 27 (1980).

\bibitem{FP80}
W. Furmanski and R. Petronzio, {\sl Singlet parton densities beyond leading order}, Phys. Lett. B {\bf 97}, 437 (1980).

\bibitem{GW}
D.J. Gross and F. Wilczek, {\sl Ultraviolet behavior of non-Abelian gauge theories}. Phys. Rev. Lett. {\bf 30}, 1343 (1973).

\bibitem{HDP}
H.D. Politzer, {\sl Reliable perturbative results for strong interactions?}, Phys. Rev. Lett. {\bf 30}, 1346 (1973).

\bibitem{KAP}
J. Kubar-Andre and F.E. Paige, {\sl Gluon corrections to the Drell-Yan model}, Phys. Rev. D {\bf 19}, 221 (1979).

\bibitem{AEM}
G. Altarelli, R.K. Ellis, and G. Martinelli, {\sl Large perturbative corrections to the Drell-Yan process in QCD}, Nucl. Phys. B {\bf 157}, 461 (1979).

\bibitem{HvN}
B. Humpert and W.L. van Neerven, {\sl Infrared and mass regularization in AF field theories (II). QCD}, Nucl. Phys. B {\bf 184}, 225 (1981).

\bibitem{DSZ}
A. Djouadi, M. Spira, and P.M. Zerwas, {\sl Production of Higgs bosons in proton colliders. QCD corrections},
Phys. Lett. B {\bf 264}, 440 (1991).

\bibitem{SD}
S. Dawson, {\sl Radiative corrections to Higgs boson production}, Nucl. Phys. B {\bf 359}, 283 (1991).

\bibitem{V2qq}
T. Matsuura, S.C. van der Marck, and W.L. van Neerven, {\sl The calculation of the second order soft and virtual contributions to the Drell-Yan cross section}, Nucl. Phys. B {\bf 319}, 570 (1989).

\bibitem{RVH}
R.V. Harlander, {\sl Virtual corrections to $gg \to H$ to two loops in the heavy top limit}, Phys. Lett. B {\bf 492}, 74 (2000) [arXiv:hep-ph/0007289].

\bibitem{WEC}
W.E. Caswell, {\sl Asymptotic behavior of non-Abelian gauge theories to two-loop order}, Phys. Rev. Lett. {\bf 33}, 244 (1974).

\bibitem{DRTJ}
D.R.T. Jones, {\sl Two-loop diagrams in Yang-Mills theory}, Nucl. Phys. B {\bf 75}, 531 (1974).

\bibitem{ET79}
E. Egorian, O.V. Tarasov, {\sl Renormalization of Quantum Chromodynamics in the two-loop approximation  in arbitrary gauge}, Teor. Mat. Fiz. {\bf 41}, 26 (1979), Theor. Math. Phys. {\bf 41}, 863 (1979).

\bibitem{Kanemura:2004mg}
S.~Kanemura, Y.~Okada, E.~Senaha, and C.P.~Yuan, {\sl Higgs coupling constants as a probe of new physics},
Phys. Rev. D {\bf 70}, 115002 (2004) [arXiv:hep-ph/0408364].

\bibitem{Misiak:2017bgg}
M.~Misiak and M.~Steinhauser,
{\sl Weak radiative decays of the $B$ meson and bounds on $M_{H^\pm }$ in the Two-Higgs-Doublet Model}, Eur. Phys. J. C {\bf 77}, 201 (2017)
[arXiv:1702.04571].

\bibitem{MG5}
J. Alwall {\it et al.}, {\sl The automated computation of tree-level and next-to-leading order differential cross sections, and their matching to parton shower simulations}, JHEP {\bf 07}, 079 (2014) [arXiv:1405.0301].

\bibitem{WuModel}
{\sl https://github.com/ycwu1030/2HDM\_FR}

\bibitem{MSHT20}
J. McGowan, T. Cridge, L.A. Harland-Lang, and R.S. Thorne, {\sl Approximate N$^3$LO parton distribution functions with theoretical uncertainties: MSHT20aN$^3$LO PDFs}, Eur. Phys. J. C {\bf 83}, 185 (2023) [arXiv:2207.04739].

\end{thebibliography}
\end{document}